\documentclass{article}
\usepackage{epsfig}
\def\S12{\mathrm{S}_{1,2}}

\newcommand{\halb}{\mbox{$\frac{1}{2}$}}

\newcommand{\IM}{[1]}

\newcounter{allequation}

\begin{document}
\hyphenation{counterterm underestimate dependent respectively Singapore}

\title{
  {\em Electroweak Corrections to the Top Quark Decay}\\[1cm]} 
\author{S. M. Oliveira${}^{1,b}$, 
  L. Br\"ucher${}^{2,b}$,
  R. Santos${}^{3,b,c}$,
  A. Barroso${}^{4,a,b}$\\[0.5cm]
  \small{${}^{a}$Dept. de F\'\i sica, Faculdade de Ci\^encias, Universidade de Lisboa}\\
  \small{  Campo Grande, C1, 1749-016 Lisboa, Portugal}\\[0.3cm]
  \small{${}^{b}$Centro de F\'\i sica Te\'orica e Computacional, Universidade de Lisboa,}\\
  \small{Av. Prof. Gama Pinto 2, 1649-003 Lisboa, Portugal}\\[0.3cm]
  \small{${}^{c}$Instituto Superior de Transportes, Campus Universit\'ario}\\
  \small{R. D. Afonso Henriques, 2330 Entroncamento, Portugal}\\[1cm]}

\date{\today} \maketitle

\begin{abstract}
We have calculated the one-loop electroweak corrections to the decay 
$t\rightarrow bW^+$, including the counterterm for the CKM matrix
elements $V_{tb}$. Previous calculations used an incorrect $\delta
V_{tb}$ that led to a gauge dependent amplitude. However, since the
contribution stemming from $\delta V_{tb}$ is small,  those
calculations only underestimate the width by roughly one part in
$10^5$.
\end{abstract}
\vspace*{3cm}

\begin{flushleft}
  PACS number(s): 12.15.Lk, 11.10.Gh, 12.15.Ji
\end{flushleft}

\footnotetext[1]{e-mail: smo@cii.fc.ul.pt}
\footnotetext[2]{present address IBM Global Services Gustav-Heinemann-Ufer 120-122, 50968 K\"oln, Germany}
\footnotetext[3]{e-mail: rsantos@cii.fc.ul.pt}
\footnotetext[4]{e-mail: barroso@cii.fc.ul.pt}

\thispagestyle{empty}

\newpage


Due to its large mass, $m_t=174.3\pm5.1$ $GeV/c^2$ \cite{Groom:2000in}, the top
quark, t, decays almost exclusively into a bottom quark, b, and a
W-boson. This two-body channel, which is not available to the other
quarks, makes the top quark singular. In fact, it is the only known
quark where the weak decay takes place before the strong hadronization 
process. Hence, contrary to the other hadronic weak processes, one
can calculate the width for the transition $t\rightarrow bW^+$ without 
being involved with the non-perturbative aspects of QCD. Because of
this advantage this process will be a good testing ground for models
beyond the Standard Model (SM). On the other hand, within the SM, the
experimental measurement of the decay rate $\Gamma\left(t\rightarrow
bW^+\right)$ gives a direct measurement of the $V_{tb}$ element of the 
Cabbibo-Kobayashi-Maskawa (CKM) matrix \cite{CKM}.

Presently, the direct observation of the top quark at the Tevatron
\cite{Tollefson:1998fv} implies that $V_{tb}$ is known with a $30\%$ error. 
The Particle Data Book \cite{Groom:2000in} gives $V_{tb}$ with a smaller error, 
but using the CKM unitarity conditions.
At the CERN LHC, with $10^7$ or $10^8$ top pairs per year, one expects to
extract $V_{tb}$ with an error of the order $10\%$ \cite{unknown:1999fr}. It
is then desirable to calculate the top width with a few percent
precision.

At tree-level the $t\rightarrow bW^+$ width, $\Gamma_0$, is
\begin{equation}\label{eq:ampG0}
\Gamma_0=\frac{\alpha}{8\sin^2\theta_W}|V_{tb}|^2\frac{\left[m_t^2-(m_W+m_b)^2\right]^{1/2}\left[m_t^2-(m_W-m_b)^2\right]^{1/2}}{m_t}\left[\frac{m_t^2+m_b^2}{2m_t^2}+\frac{(m_t^2-m_b^2)^2}{2m_t^2m_W^2}-\frac{m_W^2}{m_t^2}\right],
\end{equation}
where $\alpha=1/137.03599$ is the fine-structure constant and
$\theta_W$ is the Weinberg angle
$\left(\cos\theta_W=\frac{m_W}{m_Z}\right)$. The main
correction to $\Gamma_0$ stems from the one-loop gluon correction to
the weak vertex. This ${\cal O}(\alpha\alpha_s)$ contribution was
first evaluated by Jezabek and K\"{u}hn  \cite{Jezabek:1989iv}, and
later confirmed by Denner and Sack(DS) \cite{Denner:1991ns} and Eilam et
al. \cite{Eilam:1991iz}. Recently, a similar result was
obtained \cite{Ghinculov:2000nx} applying the optical theorem to the
two-body self-energy of the top quark. At order $(\alpha\alpha_s^2)$
there are two calculations. Czarnecki and Melnikov
 \cite{Czarnecki:1999qc} evaluated the two-loop vertex diagram for
$t\rightarrow bW^+$ using the $m_W=0$ approximation, while Chetyrkin
et al. \cite{Chetyrkin:1999ju} expanded the imaginary part of the
three-loop self-energy as a series in $q^2/m_t^2$. A recent approach
to the same problem by Chinculov and Yao \cite{Ghinculov:2000nx} uses a 
combination of analytical and numerical methods to evaluate the
general massive two-loop Feynman diagrams. The electroweak
corrections of order $\alpha^2$ were only evaluated in
refs. \cite{Denner:1991ns} and  \cite{Eilam:1991iz}. However, as Gambino,
Grassi and Madricardo (GGM) \cite{Gamb}  have pointed out, in these papers the
renormalization of $V_{tb}$ was done in such a way that the final
result was gauge dependent. 
Recently we \cite{Barroso:2000is} have considered the renormalization
of the CKM matrix, $V_{Ij}$, in the generic linear $R_\xi$ gauge. We
have confirmed that the DS \cite{Denner:1991ns}
renormalization prescription leads to a gauge dependent amplitude and
we have solved the problem introducing a condition to fix
$\delta V_{Ij}$ different from the one proposed by GGM \cite{Gamb}. Despite the fact that DS \cite{Denner:1991ns} have used 
a gauge dependent $\delta V_{Ij}$ their numerical values for the $W$
partial decay widths are essentially correct. In fact, the $\delta
V_{Ij}$ contribution is negligible. This was confirmed by
Kniehl et al. \cite{Kniehl:2000rb} using the GGM prescription. Clearly, it is in the top decay process that a wrong 
 $\delta V_{Ij}$ would induce the largest difference.
 In view of this situation we think that it is
worthwhile to present, in this note, the correct result for the
electroweak one-loop top decay. We will compare our renormalization
scheme \cite{Barroso:2000is} with the one proposed by GGM \cite{Gamb}.

Denoting by $p$ and $q$ the four-momenta of the incoming top quark and
the outgoing $W^+$, respectively, the tree level decay amplitude
$T_0$ is:
\begin{equation}\label{eq:ampT0}
  T_0=V_{tb}A_L,
\end{equation}
with
\begin{equation}\label{eq:AL}
  A_L=\frac{g}{\sqrt{2}}\bar{u}(p-q)\slash{\varepsilon}\gamma_Lu(p),
\end{equation}
where $\varepsilon^\mu$ is the polarization vector and, as usual,
$\gamma_L=(1-\gamma_5)/2$. The one-loop amplitude $T_1$ can be written
in terms of four independent form factors, $F_L$, $F_R$, $G_L$ and
$G_R$, each one associated with a given Lorentz structure for the
spinors. $F_L$ is associated with $A_L$ and $F_R$ with $A_R$ which is
given by eq.(\ref{eq:AL}) replacing $\gamma_L$ by
$\gamma_R$. Similarly, $G_L$ and $G_R$ are multiplied by $B_L$ and
$B_R$, respectively, given by:
\begin{equation}\label{eq:B_LR}
  B_{L,R}=\frac{g}{\sqrt{2}}\bar{u}(p-q)\frac{\varepsilon.p}{m_W}\gamma_{L,R}u(p).
\end{equation}
Besides the form factors, $T_1$ also depends on the counterterms. The
final result is:
\begin{eqnarray}
  \label{eq:T1}
  T_1 & = & A_L \left [\, V_{tb}\,\left(\,F_L +\frac{\delta g}{g} + \halb
  \delta Z_W + \halb \delta Z_{tt}^{L*} + \halb \delta Z_{bb}^L\right)
   \,+\,\sum_{I \neq t} \halb \delta Z_{It}^{L*}
  V_{Ib} \,+\,\sum_{j\neq b} V_{tj} \halb
  \delta Z_{jb}^{L} \,+\, \delta V_{tb} \,\right] \nonumber \\
  & & \,+\, V_{tb} \, \left [ \, A_R F_R + B_L G_L + B_R G_R \,\right]
  \enskip .
\end{eqnarray}
A detailed discussion of the counterterms can be found in our previous 
work  \cite{Barroso:2000is} and so there is no need to repeat it
here. In particular, we have shown  \cite{Barroso:2000is} that one
obtains a finite and gauge invariant $T_1$ with the $V_{tb}$
counterterm, $\delta V_{tb}$, given by:
\begin{eqnarray}
  \label{eq27}
  \delta V_{tb} & = &    -\halb \sum_{I\neq t} \delta Z^{L*}_{It} V_{Ib}
   -\halb \sum_{j\neq b} V_{tj} \delta Z^{L}_{jb} 
   -\halb V_{tb}\left[ 
    \delta Z_{tt}^{L*} -\delta Z_{tt \IM}^{L*} 
   +\delta Z_{bb}^{L} -\delta Z_{bb \IM}^{L} \right] ,
\end{eqnarray}
where $\delta Z^{L}_{II^\prime}$ and  $\delta Z^{L}_{jj^\prime}$ are
the up and the down left-handed quark wave functions renormalization
constants, respectively. A $\delta Z$ with the subscript $\IM$ means that in
its evaluation the CKM matrix was replaced by the identity matrix.

Let us stress that the only difference between our calculation and
the previous ones \cite{Denner:1991ns,Eilam:1991iz} is entirely due to
a different choice of $\delta V_{tb}$. Unfortunately, the choice made
by DS  \cite{Denner:1991ns} is not physically
acceptable. However, as we will see, $\delta V_{tb}$ gives a rather small
contribution. Hence, the numerical result does not show any meaningful
change. Perhaps, the best way to discuss the result is to define
$\delta$ as:
\begin{equation}
  \delta =\frac{2Re[T_0T_1^+]}{|T_0|^2}.
\end{equation} 
This, in turn, means that up to ${\cal O}(\alpha^2)$ the decay
amplitude can be written as:
\begin{equation}
  \Gamma =\Gamma_0[1+\delta].
\end{equation} 

In table \ref{table1} we show the different contributions to $\delta$
arising from the individual terms of eq.(\ref{eq:T1}). In the
calculations the program packages FeynArts \cite{Kublbeck:1990xcHahn:2000kx}, FeynCalc \cite{Mertig:1991an} and LoopTools \cite{Hahn:1999yk} were used.
Notice that,
with our renormalization prescription for $\delta V_{tb}$, all
contributions from the off-diagonal quark wave-functions
renormalization constants are canceled and one simply needs to
evaluate $\delta Z^{L*}_{tt \IM}$ and $\delta Z^{L}_{bb \IM}$. They,
together with the other counterterms give a large positive $\delta$
$(23.66\%)$ which is then reduced to $4.46\%$ with the negative
contribution of $F_L$ $(-18.75\%)$ and $G_R$ $(-0.44\%)$. The other
form factors give negligible contributions.
It is interesting to see the difference when we
follow the CKM renormalization prescription given by GGM \cite{Gamb}. The calculation is slightly more complicated: the
off-diagonal terms proportional to $\delta Z^{L*}_{It}$ and $\delta
Z^{L}_{jb}$ have to be included; the diagonal terms $\delta
Z^{L*}_{tt}$ and $\delta Z^{L}_{bb}$ have to be calculated without the 
approximation of replacing the CKM matrix by the unit matrix; and
finally one ought to add $\delta V_{tb}^G$ given by:
\begin{eqnarray}
\delta V_{tb}^G=\frac{1}{2}\left[\sum\limits_{I} \delta
    {\cal Z}_{tI}^{L,A} V_{Ib}- 
  \sum\limits_{j} V_{tj}\delta {\cal Z}_{jb}^{L,A} \right], 
\end{eqnarray}
where the $\delta {\cal Z}_{ij}^{L,A}$ are ``special'' anti-hermitian wave 
function renormalization constants fixed in terms of the quark self-energies
at $q^2=0$, namely,
\begin{eqnarray}
\delta{\cal
  Z}_{ij}^{L,A}&=&\frac{m_i^2+m_j^2}{m_i^2-m_j^2}\left[
{\Sigma}_{ij}^L(0)+2{\Sigma}_{ij}^S(0)\right] . 
\end{eqnarray}
\begin{table}[htbp]
  \begin{center}
    \begin{tabular}{|c|c|}
      \hline
      Form Factors and Counterterms & Contributions to $\delta(\%)$\\
      \hline
      $F_L$&$-18.753$\\
      $F_R$&$-2\times 10^{-3}$\\
      $G_L$&$-8\times 10^{-4}$\\
      $G_R$&$-0.445$\\
      $\frac{\delta g}{g}$&$10.419$\\
      $\frac{1}{2}\delta Z_W$&$3.193$\\
      $\frac{1}{2}\delta Z_{tt[1]}^L$&$5.220$\\
      $\frac{1}{2}\delta Z_{bb[1]}^L $&$4.831$\\
      \hline
      Total&$4.46$\\
      \hline
      \hline
      $\halb \sum_{I\neq t} \delta Z^{L*}_{It} V_{Ib}$&$9\times 10^{-6}$\\
      $\frac{1}{2}\sum_{j\neq b}V_{tj}\delta Z_{jb}^L$&$-1.8\times 10^{-3}$\\
      $\frac{1}{2}\delta Z_{tt}^L-\frac{1}{2}\delta Z_{tt[1]}^L$&$-0.1\times 10^{-3}$\\
      $\frac{1}{2}\delta Z_{bb}^L-\frac{1}{2}\delta Z_{bb[1]}^L $&$-5.3\times 10^{-3}$\\
      $\delta V_{tb}^G$&$6.4\times 10^{-3}$\\
      \hline
      Total&$-0.8\times 10^{-3}$\\
      \hline
    \end{tabular}
    \caption{Contributions to $\delta$ from the individual terms in
      eq.(\ref{eq:T1}) evaluated at $m_t=174.3$ $GeV/c^2$ and
      $m_H=114$ $GeV/c^2$. The lower part lists the additional
      contributions needed if the GGM \cite{Gamb} renormalization
      prescription is used.}
    \label{table1}
  \end{center}
\end{table}
For the sake of completeness we have also listed in table 1 the numerical values of these additional contributions. They are all extremely small which means that $\delta$ is practically the same in both renormalizations schemes. 

\begin{figure}[htbp]   
  \begin{center}
    \epsfig{file=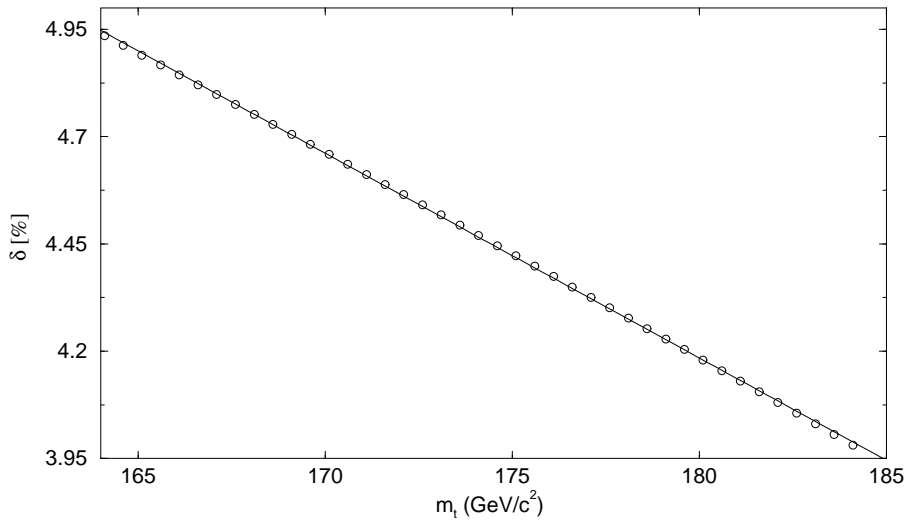,angle=-90, width=12cm}
    \caption{$\delta$ as a function of the top mass.}
    \label{fig_mtop}
  \end{center}
\end{figure}

\begin{figure}[htbp]
  \begin{center}
    \epsfig{file=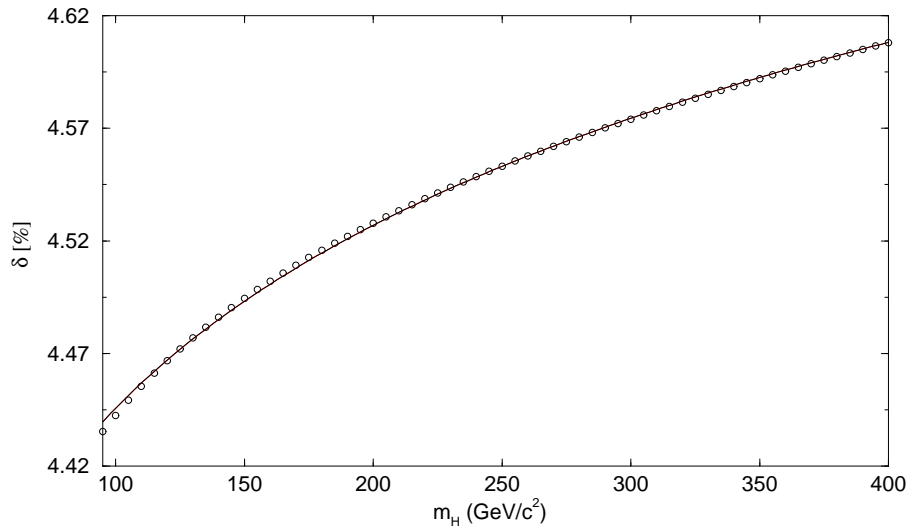,angle=-90, width=12cm}
    \caption{$\delta$ as a function of the Higgs mass.}
    \label{fig_mH}
  \end{center}
\end{figure}

Certainly, the uncertainty introduced in the calculation by the error in the top quark mass is far more important. To illustrate this remark and to avoid the need to repeat this calculation in the future we have done it varying $m_t$ in the two-sigma interval around the present experimental mean value. We have found, that within this interval the value of $\delta$ can be very well reproduced by the linear fit:
\begin{equation}
  \delta =\left[12.7715-0.0477\frac{m_t}{GeV/c^2}\right]\times 10^{-2} .
\end{equation} 
Figure \ref{fig_mtop} shows the quality of this fit. 
Another parameter that enters the calculation is the Higgs mass
$m_H$. In the results given in table \ref{table1} and in
fig. \ref{fig_mtop} we have used rather arbitrarily $m_H=114$
$GeV/c^2$. As it is well known $\delta$ depends logarithmically on $m_H$.
Again for $m_t=174.3$ $GeV/c^2$ and for $100$ $GeV/c^2\leq m_H\leq
400$ $GeV/c^2$, $\delta$ could be fitted with the following expression:
\begin{equation}
  \delta =\left[4.4457+0.1172\,\ln\frac{m_H}{100\ GeV/c^2}\right]\times 10^{-2} .
\end{equation} 
In figure \ref{fig_mH} we show the result and the fitted curve.

We would like to summarize our conclusions as follows:
\begin{enumerate}
\item Using our  \cite{Barroso:2000is} prescription for the renormalization of the CKM matrix elements we have calculated the electroweak radiative corrections to the decay width $t\rightarrow bW^+$;
\item For $m_t=174.3GeV/c^2$ and $m_H=114$ $GeV/c^2$, the correction
  is $\delta=4.46\%$. This increases the tree level value of $\Gamma$
  from $1.4625$ $GeV/c^2$ to $1.5277$ $GeV/c^2$;
\item We have checked that an alternative renormalization prescription advocated by GGM \cite{Gamb} gives a width that differs from ours by less than one part in $10^{5}$;
\item The contribution to $\delta$ stemming from the $\delta V_{tb}$
  counterterm is rather small. It is $7.2\times 10^{-3}\, \%$ versus
    $6.4\times 10^{-3}\, \%$ in the GGM \cite{Gamb} scheme, while 
   the old DS \cite{Denner:1991ns}  $\delta V_{tb}$ counterterm would
   have given $6.6\times 10^{-3}\, \%$.

\end{enumerate}
\newpage
{\bf Acknowledgement}

This work is supported
by Funda\c{c}\~ao para a Ci\^encia e Tecnologia under contract
No. CERN/P/FIS/15183/99. L.B. and S.O. are supported by FCT
under contracts No. BPD.16372. and BM-20736/99.



\end{document}